\documentstyle[aps,epsf,slashbox]{revtex}

\tighten
\newcommand{\beq}{\begin{equation}}
\newcommand{\eeq}{\end{equation}}
\newcommand{\ba}{\begin{eqnarray}}
\newcommand{\ea}{\end{eqnarray}}
\newcommand{\nn}{\nonumber}

\newcommand{\bm}[1]{\bbox{#1}}

\begin{document}
 
\draft
\title{
\begin{flushright}
\begin{minipage}{4 cm}
\small
hep-ph/0204207\\
NIKHEF/2002-001
\end{minipage}
\end{flushright}
Exploring the QED vacuum with laser interferometers}
\author{Dani\"el Boer$^1$ and Jan-Willem van Holten$^{1,2}$}
\address{\mbox{}\\
$^1$ 
Division of Physics and Astronomy, Vrije Universiteit, De Boelelaan 1081\\
NL-1081 HV Amsterdam, The Netherlands\\
\mbox{}\\
$^2$ NIKHEF, P.O. Box 41882, NL-1009~DB Amsterdam, The Netherlands
}

\maketitle
\begin{center}\today \end{center}

\begin{abstract}
It is demonstrated that the nonlinear, and as yet unobserved, QED effect of 
slowing down light by application of a strong magnetic field may be 
observable with large laser interferometers like for instance LIGO or GEO600. 
\end{abstract}


The nonlinear properties of the QED vacuum in the presence of an external
electromagnetic field have been studied theoretically for more than 65 years. 
The famous Euler-Heisenberg lagrangian \cite{EH} is a low-energy effective 
lagrangian describing such physics of (multiple) photon interactions. 
The latter interactions arise effectively due to interactions with the 
electron field which is integrated out. The lowest order consequences of 
these effective photon interactions are light-by-light scattering, two-photon 
splitting and the effects on photon propagation, such as birefringence and
dichroism. Thus far these effects have 
not been observed directly yet, since they are extremely small. For example, 
the light-by-light scattering cross section is proportional to 
$\alpha^4 \omega^6/m^8$, where the center of mass photon frequency 
$\omega$ is assumed to be very small compared to the electron mass $m$.

Another QED vacuum effect, the Schwinger mechanism of $e^+e^-$ pair 
production in an electric field, which requires very large fields $|eE| 
\sim m$, has only recently been observed \cite{Burke}. 

Here we will investigate the effect of a strong magnetic field on photon 
propagation, and argue that this effect 
may be observable by exploiting one of the 
large laser interferometers that are designed to try to detect gravitational
waves. We start by observing that light propagating in a magnetic field no
longer propagates with velocity $v=c$. Using the Euler-Heisenberg lagrangian 
(for instance given in SI units)
\ba
{\cal L} & = & - \frac{c^2 \epsilon_0}{4} F_{\mu\nu}F^{\mu\nu} + 
\frac{\alpha^2 \hbar^3 \epsilon_0^2}{90 m^4 c} \left[
\left(F_{\mu\nu}F^{\mu\nu}\right)^2 + \frac{7}{4} 
\left(F_{\mu\nu}\tilde{F}^{\mu\nu}\right)^2\right] \nn\\[2 mm]
& = & \frac{\epsilon_0}{2} \left(\bm{E}^2- c^2\bm{B}^2\right) + 
\frac{2\alpha^2 \hbar^3 \epsilon_0^2}{45 m^4 c^5} \left[
\left(\bm{E}^2-c^2\bm{B}^2\right)^2 + 7 c^2 
\left(\bm{E}\cdot\bm{B}\right)^2\right], 
\label{EH}
\ea
the first order correction to the velocity has 
been derived several times already \cite{Adler,Birula,Heyl,Gies,Lorenci} and 
reads:
\beq
\frac{v}{c} = 1-a \; \frac{\alpha^2 \hbar^3 \epsilon_0}{45 m^4 c^3} \; B^2 \; 
\sin^2 \theta_B,
\label{vB}
\eeq
where $\theta_B$ is the angle between the 
direction of the photon propagation and the magnetic field $B$. These two
directions span a plane, and the constant  
$a$ is either $8$ or $14$ for the so-called $\parallel$ or $\perp$ modes of 
the photon polarization, that are parallel or perpendicular to 
that plane. 

This first
order correction given in Eq.\ (\ref{vB}) is sufficient in case the magnetic 
field is small compared to what is usually called the critical magnetic 
field $B_{cr} \equiv m^2 c^2/e\hbar \approx 4.4 \cdot 10^{9}$ T. 
Another relevant remark is that 
Eq.\ (\ref{vB}) holds independently of the frequency of the photon. 

Like in Ref.\ \cite{Birula} we will express the results in terms of the 
quantity $\kappa$:
\beq
\kappa = \frac{2 \alpha^2 \hbar^3}{45 m^4 c^5} \approx 2.7 \cdot 10^{-40}
\frac{\text{m}^3}{\text{GeV}} 
\eeq
or 
\beq
c^2 \epsilon_0 \kappa \approx 1.3 \cdot 10^{-24}\, \text{T}^{-2}.
\eeq
Hence, in Heaviside-Lorentz units (in which $\epsilon_0=1$ and 
$\alpha = e^2 /4\pi \hbar c$) one finds $\kappa \approx 1.3 
\cdot 10^{-24} (\text{Tc})^{-2}$, which is actually three orders of magnitude
smaller than the value quoted in 
Eq.\ (29) of Ref.\ \cite{Birula}. In Gaussian units (where
$\epsilon_0=1/4\pi$ and $\alpha = e^2
/\hbar c$) one arrives at $\kappa \approx 1.7 \cdot 10^{-23}$ (Tc)$^{-2}$. 
Of course, the value of the correction in Eq.\ (\ref{vB}) stays the same in 
both units. In terms of the critical magnetic field $B_{cr}$:
\beq
\frac{v}{c} = 1-a \; \frac{\alpha}{180 \pi} \; \frac{B^2}{B_{cr}^2} \; 
\sin^2 \theta_B.
\label{vB2}
\eeq

For $\theta_B=90^\circ$ Eq.\ (\ref{vB}) becomes simply $v/c = 1-a
c^2 \epsilon_0 \kappa B^2/2$, 
which for $a=14$ yields $v/c \approx 1- 10^{-23}\,\text{T}^{-2}\, B^2$. 

In order to observe such effect of $v<c$ due to the presence of a strong 
magnetic field, one could think of trying to observe light propagation 
through the magnetic field of a neutron star, which can be of the order of 
$10^8$ Tesla. In theory, this could be done with 
a method analogous to the measurement of the speed of light using a moon of
Jupiter, but now applied to binary stars. However, this does not seem 
feasible in practice.  

Here we want to advocate a different approach to the problem, namely not
by exploiting extremely large magnetic fields, but by using a large laser 
interferometer, such as LIGO or GEO600 (both soon operational), 
in order to measure propagation of 
light very accurately. If one were to apply a homogeneous strong magnetic
field over a stretch of one of the two legs of the interferometer
(with $B$ perpendicular to that leg), then one affects the velocity of the
laser light slightly each time it traverses the magnetic field. This would
result in a phase shift. In order to demonstrate the feasibility of
this approach, we consider the example of LIGO \cite{LIGO}. 
We will call the length of 
the legs $x$, therefore $x$ is taken to be 4 km (for GEO600: $x=600$ m), 
and we assume the design 
sensitivity $\delta x /x$, which is of the order of $10^{-21}$. 

In the case of a gravitational wave passing the interferometer, $\delta x$ is
a real variation in the length of a leg (with alternating sign). 
In the case of an applied magnetic
field $\delta x$ is just the difference in distance propagated by light which
travels with $c$ over the whole distance $x$ compared to light that
travels part of $x$, i.e.\ a stretch $x'$, with a slightly lower velocity 
(which has
a fixed sign and is therefore a cumulative effect). 
One finds in that case that 
\beq
\frac{v}{c} \approx  1 - \frac{\delta x}{x'}.
\eeq
Combining this with Eq.\ (\ref{vB}) yields
\beq
\frac{\delta x}{x} = \frac{x'}{x} \frac{a c^2 \epsilon_0 \kappa}{2} B^2 \; \; 
\stackrel{a=14}{\approx} \; \; \frac{x^{\prime}}{x} \times 10^{-23}\, 
\left( \frac{B}{1 \, \mbox{T}} \right)^2
\eeq 
which for LIGO with for instance $x'=4$ m (for GEO600 one arrives at the same
numbers if $x' = 60$ cm), would imply a very large magnetic field 
$B^2 = 10^5$ T$^2$. Conversely, if for
example, one would apply the magnetic field of 1 or 10 Tesla over $x'=4$ m,
then this would require $\delta x/x$ to be $10^{-26}$ or $10^{-24}$,
respectively. This is not expected to be
reached any time soon in the case of laser interferometer gravitational wave 
detectors. 

Another important observation is that the number of
round trips that the light travels in the cavities of such an 
interferometer, 
is matched to the period of half a cycle of a gravitational wave. For
instance, for a
100 Hz gravitational wave this means almost 200 round trips inside the LIGO
detector in the 
duration of $5\cdot 10^{-3}$ seconds and the phase shift accuracy
corresponding to $\delta x/x = 10^{-21}$ is then 
roughly $\delta \Phi \approx 200 \cdot 4 \pi \delta x /\lambda \approx 
2 \cdot 10^{-8}$ for $\lambda \approx 0.5 \mu$m; for more details cf.\  
Refs.\ \cite{LIGO,Saulson,Thorne}. 
We will use the phase shift $\delta \Phi_{\text{thr}} = 2
\cdot 10^{-8}$ as the detection threshold for LIGO in our example. Note that
strictly speaking the wavelength of the light traversing the magnetic field is
slightly lower than the initial $\lambda$ 
(the frequency remains unchanged), but this amounts to a
higher order (in $\delta x$) effect in the calculation of $\delta \Phi$ and
can be safely neglected. 

Now in order to measure the effect of the
QED vacuum on the propagation of light in a magnetic background field, one 
can simply let the light run more cycles in the interferometer. This will of
course affect the noise analysis and photon loss due to scattering and
absorption will become a more important issue. But if one simply 
takes $\delta x/x = 10^{-21}$ and multiplies 
with a factor $n$ of required round trips compared to the gravitational wave 
detection scenario, then one gets in the above example 
$n B^2= 10^5$~T$^2$ in order to obtain the threshold phase shift. 
Hence, for $B=10$ T one finds the required number of
round trips is $n=10^3$ times the number of round trips to detect a
gravitational wave. This means a build-up time of the phase shift 
of 5 seconds, which seems feasible. 

For convenience, we include a table which
shows the total number $N$ of round trips for fixed 
$\delta \Phi_{\text{thr}} = 2 \cdot 10^{-8}$, but for different 
values of the magnetic field and $x'$ (the results are independent of the
length $x$ of the interferometer for
given $\delta \Phi_{\text{thr}}$). Note that the number of round trips $N$ 
is directly related to the finesse of the Fabry-Perot cavity that is used. 

\begin{center}
\begin{minipage}{17 cm}
\begin{table}[htb]
\begin{center}
\begin{minipage}{7 cm}
\begin{tabular}{|c||*{3}{c|}}
\backslashbox{$B^2$}{$x'$}
&\makebox[3em]{$1$ m}&\makebox[3em]{$2$ m}&\makebox[3em]{$4$ m}
\\[1 mm]\hline\hline\\[-3 mm]
1 \mbox{T}$^2$ & $8 \cdot 10^{7}$ & $4 \cdot 10^{7}$ & $2 \cdot 10^{7}$\\[1 mm]\hline\\[-3 mm]
10 \mbox{T}$^2$ & $8 \cdot 10^{6}$ & $4 \cdot 10^{6}$ & $2 \cdot 10^{6}$\\[1 mm]\hline\\[-3 mm]
100 \mbox{T}$^2$ & $8 \cdot 10^{5}$ & $4 \cdot 10^{5}$ & $2 \cdot 10^{5}$\\[1 mm]
\end{tabular}
\end{minipage}\\[3 mm]
\caption{Examples of the total number of round trips $N$ for several values 
of the $B$ field and $x'$, for 
$\delta \Phi_{\text{thr}} = 2 \cdot 10^{-8}$}
\end{center}
\end{table}
\end{minipage}
\end{center}

We add a few more remarks. 
One might like to exploit the maximal phase difference
that will be acquired over time,
but in order to reach a phase shift $\delta \Phi= \pi$ in our example
one needs to let the light run for approximately 25 years, which is not a
realistic goal. Moreover, since $v_\perp \neq v_\parallel$,
different polarization states would not have this phase shift at the
same time. 

It may be good to emphasize that for a measurement of vacuum birefringence
(the fact that $n_\parallel \neq n_\perp$ and hence $v_\perp \neq 
v_\parallel$) one needs to polarize the light to be able to select different
polarization modes. But in order to demonstrate the dispersive effect of 
slowing down light by application of a strong magnetic field, one can 
simply deal with an average of polarization states. Demonstrating the 
intensity variations with $B^2$ seems to be the best way to establish this 
effect. 
 
We would like to refer to other proposals given in the literature. 
Refs.\ \cite{Iacopini,Cantatore} propose to measure vacuum birefringence 
from the (essentially increasing) phase difference between the
$\parallel$ and $\perp$ modes of a laser beam passing through a magnetic
field (no use is made of an interferometer). 
This forms the basis for the PVLAS experiment \cite{Bakalov} 
and requires control over the initial polarization state of the light.
As explained above this is not necessary for the dispersion measurement 
suggested in this note, which (ideally) 
aims at measuring the absolute velocity 
decrease. 

There is also a proposal \cite{Partovi} that tries to exploit a laser  
interferometer, but requires an additional laser beam crossing alternatingly
the laser beams of the interferometer. The measurement is thus based on 
light-by-light scattering. This is quite different from the present proposal.
 
In conclusion, with the advent of large laser interferometers with extreme
precision ($\delta x/x \sim 10^{-21}$ and better) and the feasibility of
creating magnetic fields above 1 Tesla, another possibility to observe 
photon-photon interactions due to QED vacuum effects may be realized. 

\acknowledgments 
D.B.\ would like to thank Jens O. Andersen for a very helpful discussion. The 
research of D.B.\ has been made possible by financial support from the Royal 
Netherlands Academy of Arts and Sciences.

\end{document}